1  FLARE-ASSOCIATED TYPE III RADIO BURSTS AND DYNAMICS OF THE

2  EUV JET FROM SDO/AIA AND RHESSI OBSERVATIONS

4  Naihwa Chen[1], Wing-Huen Ip[1,2], and Davina Innes[3]

6  [1] Graduate Institute of Astronomy, National Central University, Jhongli 32001, Taiwan,

7  R.O.C.

8  d949001@astro.ncu.edu.tw

9  [2] Institute of Space Sciences, National Central Univeristy, Jhongli 320001, Taiwan,

10  R.O.C.

11  wingip@astro.ncu.edu.tw

12  [3] Max-Planck-Institut fuer Sonnensystemforschung, 37191, Katlenburg-Lindau,

13  Germany

14  innes@mps.mpg.de

**Abstract**

We present a detailed description of the interrelation between the Type III radio bursts and energetic phenomena associated with the flare activities in Active region AR 11158 at 07:58 UT on 2011, Feb. 15. The timing of the Type-III radio burst measured by the radio wave experiment on the Wind/WAVE and an array of ground-based radio telescopes, coincided with an EUV jet and hard X-ray emission observed by SDO/AIA and RHESSI., respectively. There is clear evidence that the EUV jet shares the same source region as the hard X-ray emission. The temperature of the jet, as determined by multiwavelength measurements of AIA, suggests that type III emission is associated with hot, 7 MK, plasma at the jet's footpoint.

Subject heading: Sun: flares --- Sun: X-rays ---Sun: type-III radio bursts

## 1. Introduction

It is generally believed that energetic activity like solar flares in the solar atmosphere is driven by magnetic reconnection in the vicinity of active regions (ARs). Formation of current sheets due to the emergence of magnetic flux near the boundaries of sunspots could be conducive to reconnection which is characterized by particle acceleration and X-ray jets (Shibata et al. 1992; Shimojo et al. 1996; 1997; 1998). The generation of an electron beam in connection with particle acceleration near the reconnection site leads to a number of impulsive physical phenomena which can in turn be used as diagnostic tools for solar flares. These include the injection of near-relativistic elections into interplanetary space (Lin 1985; Kahler et al. 2007; Klassen et al. 2011) and the production of Type-III radio bursts with frequencies from a few tens to thousands of kHz by the streaming of low energy electrons ($< 30$ keV) upward through the corona (Suzuki et al. 1985). These two types of electron signatures have been shown to display very close correlation in time by comparing the radio wave measurements by the SWAVES (Bougeret et al. 2008) and the solar electron and proton telescope (SEPT) observations (Müller-Mellin et al. 2008) on the STEREO spacecraft, and allowing for the travel time of the energetic electrons along the interplanetary magnetic field line (Klassen et al. 2011).

The association of Type-III radio bursts with EUV jets was studied by a number of authors (Aurass et al. 1994; Kundu et al. 1994; Kundu et al. 1995; Raulin et al. 1996; Kundu et al. 2001; Pick et al. 2006; Wang et al. 2006; Nitta et al. 2008; Krucker et al. 2011). Taking advantage of the multi-wavelength EUV imaging observations of the Solar Dynamic Orbiter (SDO), Innes et al. (2011) recently demonstrated further that Type III radio bursts detected on August 3, 2010, preceded by about 30s onset of narrow EUV jets detected in 211 Å and 304 Å in an active region. The correlation of the Type-III radio waves with impulsive hard X-ray (HXR) emission was investigated by Benz et al. (2005a) and Dąbrowski and Benz (2009) by using the Reuven Ramaty High Energy Solar Spectroscopic Imager (RHESSI) (Lin et al. 2002). What is still missing is the direct comparison of all three kinds of data for jets in ARs, so that a more comprehensive picture of the reconnection process can be obtained.

Following the approach of these authors, we would like to examine the interrelations among the Type III radio bursts from the Waves/Wind experiment (Bougeret et al. 1995), Compact Astronomical Low-cost Low-frequency Instrument for Spectroscopy and Transportable Observatory (e-Callosto) (Benz et al. 2005b) and Phoenix-3 (Benz et al. 2009), the EUV jets from Atmospheric Imaging Assembly (AIA) (Lemen et al. 2012) and the HXR emission from for AR11158 between February 13 and 15, 2011. The paper is organized as follows. Section 2 will describe the instruments, observations and data analysis. Section 3 provides an analysis of the dynamical evolution and the interrelation among the Type-III radio bursts, the EUV jet feature and the HXR source region. The thermal evolution of the jet ejecta after the impulsive flare at 07:58 UT is investigated by means of multi-wavelength AIA measurements. A summary and discussion will be given in Section 4.

## 2. Instruments and Observations

AR 11158 was the first active region to appear in the rising phase of the current solar cycle 24. When it crossed the solar disk from the central meridian to the west limb between February 11 and 21, its morphology changed rapidly from simple ß- to complex B-ß gamma configuration (Tan et al. 2012). A X2.2 flare occurred at 01:33 UT on February 15. In total, 56 C-class flares and five M-class flares were produced in AR 11158 between February 13 and 18 (Beauregard et al. 2012). They give us the first opportunity to study the relation between Type III radio bursts and jets associated with flare activity using the new instruments on SDO together with an array of experiments on the GOES, RHESSI. and STEREO spacecraft. In this work, we focus our attention on a C-class flare and EUV jet which took place at 07:58 on February 15, 2011.

From SDO, we used Helioseismic and Magnetic Imager (HMI) and AIA data. The HMI images the whole solar disk in the photospheric Fe I, absorption line at 6173.3 Å with an angular resolution of 0.5"/ pxl and a cadence of 45 secs (Schou et al. 2012). AIA observes the full solar disk with an angular resolution of 0.6" and a cadence of 12 secs in a number of wavelengths which allow to probe the temperature distributions in different regions from transition region to the corona. For example, the 94 Å filter is centered on the Fe XVIII line emitted by hot plasma with $\log T \sim 6.8$, the 211Å filter reveals the regions of warm plasma in the vicinity of ARs at $\log T \sim 6.2$, and the 304 Å He II line forms in the solar chromosphere and the transition region at $\log T \sim 4.7$. The 131 Å filter cover both the hot Fe XXI line ($\log T \sim 7$) and cool Fe VIII lines ($\log T \sim 5.6$). As will be discussed later in section 3, the ratios of different line intensities can be used to deduce the plasma temperature.

The archived HXR data from RHESSI are another important component of our study. RHESSI images the full Sun in the energy range from 3 keV to 17 MeV with an angular resolution of 2.3" providing detailed information on the positions and structures of thermal and non-thermal HXR sources in different energy bands.

For radio wave measurements, we have made use of three data sets. The observations of Wind/Waves experiment cover the frequency range between 1 MHz and 14 MHz (Bougeret et al.1995). The frequency range from 45 MHz to 870 MHz with a resolution of 62.5 kHz is covered by the e-Callisto network of solar radio spectrometers. This network is composed of nine stations in different longitudes of global monitoring observations and aims at 24 hours coverage of radio emissions (Benz et al 2005b). Finally, the Phoenix-3 multi-channels radio spectrometer with a 5-meter antenna at Bleien observatory, Switzerland has the capability of measuring solar radio emission at frequency range of 1-5 GHz with a spectral resolution of 61 kHz and a time resolution of 200 ms (Benz et al. 2009).

## 3. RESULTS

### 3.1 RADIO SIGNATURES AND HXR EMISSIONS

The HXR and radio emissions are generally thought to be a signature of particle acceleration. The intensive HXR flux is often accompanied by metric and decimetric emissions in the main phase of a flare (Benz et al. 2005a). Figure 1 shows a comparison of the Type-III radio bursts as observed in different frequency ranges (from 1 MHz to >1 GHz) by the Wind/Wave experiment and ground-based observations (e-Callisto and Phoenix-3). After the first appearance at 07:58 UT, the radio signal drifted fast from a few GHz to very low frequency (~1MHz) within 2 minutes. This dynamic behavior is basically due to the outward beaming of mildly relativistic electrons from the flare site to the outer corona and interplanetary space. About 1 min before the main pulse at 07:58 UT, decimetric and metric radio emissions can be found probably indicating pre-flare electron acceleration. At the same time, some faint narrowband spikes at higher decimetric frequency shown in Phoenix-3 spectrum are well associated with the rise of HXR 30-50 keV flux. The generation of decimetric radio bursts is thought to be related to the primary energy release process. Figure 2 compares the dynamic spectra from Wind/Waves with the time profile of HXR fluxes measured by RHESSI and EUV emissions from SDO/AIA, respectively. The intense type III radio burst took place at 07:58 UT which is also the peak time of the HXR fluxes (4-10 keV) but the EUV flux peaks a few seconds later. The near simultaneous occurrences of the Type III radio bursts, EUV and HXR emissions speak for their common origin associated with the impulsive flare. It shows that the energetic electrons accelerated via reconnection can gain immediate access to the outer corona and interplanetary space.

Note that the close correlation between decimetric radio pulsations and the HXR emission has been report by Benz, et al. (2005a) and Dąbrowski and Benz (2009). The variations of HXR profiles show the formation of a secondary peak at 08:01 UT in the energy ranges of 4-10 keV and 10-30 keV, but not at the 30-50 keV channel. The absence of a Type III radio burst at the time of the second peak might mean that the generation mechanism responsible for the lower energy HXRs did not produce the required energetic electrons, although, as discussed in the next section, EUV jets were seen. Another possibility is that all accelerated electrons might be directed along closed field lines back to the Sun.

3.2 RADIO SIGNATURES AND EUV JETTING

Figure 3a shows contours of the HMI magnetogram superimposed on the AIA image taken at 211 A at the time of the peak HXR flux. There are two pairs of sunspots plus a number of smaller magnetic concentrations. Loop-like structures can be found emanating from some of these localized areas. A prominent jet feature appears on the lower left hand side of the figure. This EUV jet has its root in a small region containing two opposite magnetic polarities which might have been responsible for the reconnection. Not specifically shown in Figure 3a is the location of a region of strong HXR emission at the footpoint of the EUV jet which was situated between the two opposite polarity magnetic field concentrations. To trace the time evolution of the EUV jet and the HXR source region, contour maps of the HXR emissions in three different energy channels are aligned and overlaid on the 211 Å images at different times from 07:55 UT to 08:11 UT (as Figure 3b.). It can be seen that the HXR source region was not there at 07:55 UT but appeared suddenly one minute later at 07:56 UT. At that time, a short spike of EUV emission projected out of the HXR emission zone. The spike quickly transformed itself into a jet rooted at the non-thermal HXR source region.

At 08:03 UT we could still find a faint trace of the HXR source and some remnant of the EUV jet. By 08:11 UT everything was gone. In this eruption event, unlike those analyzed by Krucker, et al. (2011), the topology does not show two or three non-thermal sources and a thermal one, possibly because the sources were more compact and hence spatially unresolved in the HXR images (resolution ~6"). However, the EUV jet rooted in two opposite polarities with a cospatial non-thermal source indicates a possibility for reconnection and the production of fast escaping electrons. Figure 4 shows the time evolution of EUV AIA 94-, 131-, 211-, and 304 Å intensities integrated along the path of the jet during the flare eruption. The sudden appearance of the jet occurred at the same time as the impulsive phase flare at 07:58 UT. Besides the first jet, a second jet can be recognized at 08:00 UT in the AIA 211 Å image. Also shown is the Wind/WAVE spectrogram of the type-III radio burst. The intensity enhancement of the first EUV jet was associated with the radio burst, but the onset times were slightly different in the four filters. The onset of the slower jet, seen in the 211-, 304 Å filters, lagged behind the type-III burst, but the faster jet, observed in the 94- and 131 Å filters, preceded it by a few seconds. A similar short delay between radio bursts and 211 A jets was described by Innes, Cameron, and Solanki (2011). They also noted that the EUV jet started beyond the brightening in the footpoint area which is the configuration seen in this jet as well (figure 4). The second jet was easily visible in the 211-, 304 Å images but not in the other two filters. Furthermore, it was not associated with a HXR, 30-50 keV, source or a type-III radio burst. This analysis of the time evolution shows that the onset of the first jet, which was associated with a radio burst, was initially seen in 94- and 131 Å emission. The second jet was neither seen in 94- and 131-A emission nor was it associated with a radio burst. This can help us to understand the formation of EUV jets and their relation to the trigger of type-III bursts which will be discussed in the next section.

3.3 CORONAL TEMPERATURE DISTRIBUTIONS

To examine the formation of the EUV jet and its relationship with the production of the type-III burst, we looked at the plasma temperature evolution in the vicinity of AR 11158, taking advantage of the high time resolution of SDO/AIA. The coronal temperature was obtained from the differential emission measure (DEM) using data from six EUV filters (94, 131, 171, 193, 211, 335) of AIA (Aschwanden et al. 2011) and the references therein. In Figure 5, the deep blue region is the cool open-field area in the AR 11158 main loop system while the orange region, on the right, is the higher temperature closed-field area. To understand the temperature distribution and evolution in the jet structure, a black contour outlines the jet shape as observed in the 211 Å images and white thick circles indicate the HXR sources at the time of the temperature maps. For this discussion, we divide the time interval into two parts.

1. Before the type-III burst (07:55:50-07:57:26 UT)

Initially, the temperature of the low-lying emission at the site of the jet is about 1-2 MK. The temperature of the footpoint area started to rise as soon as the HXR source appeared and increased until the onset of the type-III burst. From the previous section, the footpoint brightening preceded the onset of the EUV jet, and here we find that this early brightening was accompanied by a temperature increase. There was a complex mixture of hot, warm and cool temperatures inside a cusp-like feature at the top of a low-lying loop at the base of the jet. At the same time, the radio signal drifted quickly from a few GHz to ~1 MHz (Figure 1) as the type-III-burst-associated electron beams propagated outward. This complex structure might relate to the production of type-III bursts but a more detailed analysis is required.

2. After the type-III burst (07:58:26-08:01:14 UT)

The temperature distribution along the jet was not homogeneous but segmental, although most of the jet region was at the same temperature as its surroundings. The footpoint area was always hotter while the HXR source was visible. When the second jet started (08:00 UT), the temperature of the jet rose up temporarily but without any correlated type-III burst. After the HXR sources vanished (08:05-08:10 UT), the jet shrank and its temperature slowly decreased to that of the surroundings. At the same time, the temperature of the low-lying loop also returned to its pre-radio-burst state.

In this temporal and spatial temperature analysis, we found that an early temperature increase at the jet footpoints precedes the onset of the EUV jet and type-III burst which is consistent with the early footpoint brightening in the EUV intensity-time plot. This early temperature increase might be due to flare-accelerated electrons with lower-energy (HXR source in 4-10 keV). The footpoint area sustained its high temperature until a few minutes after the type-III burst which is unfavorable for the production of the next type-III burst according to the simulation results of Li et al. (2011). This might be the reason why there was no type-III burst during the second X-ray emission peak (Figure 1). Another possibility, as discussed above, is that the reconnection and acceleration process was not strong enough to produce a non-thermal beam of electrons.

## 4. SUMMARY AND DISCUSSION

A comprehensive study has been carried out for the analysis of the interrelation between the type III radio burst, EUV jets and HXR emissions of the first active region of solar cycle 24, AR 11185, on 2011 February 15. From a consideration of the timing and spatial locations, a common origin of the type III radio bursts, the EUV jet and HXR source region can be well established. The new measurements by the SDO/AIA instrument showed how the temperature of jet plasma distributes before and after the type III radio burst occurs. Further investigations of the dynamics of such EUV jets with observations from SDO, RHESSI and space-borne and ground-based radio telescopes would bring new insights to the corresponding particle energization and reconnection process.

## 5. Acknowledgment

We thank Bernard Jackson and Ya-Hui Yang for useful information and suggestions. We also thank referees for many useful comments and suggestions. This work is supported by NSC grant: NSC 96-2752-M-008-011-PAE and by the Ministry of Education under the Aim for Top University Program, NCU.

Figure Captions

Figure 1. A comparison of the Type-III radio burst at different frequency ranges associated with AR 11158 at UT 07:58 on February 15, 2011. From top to bottom: Phoenix 3, Bleien, Ooty, Wind/Waves.

Figure 2. A comparison of the Type-III radio burst at 07:58 UT on February 15, 2011, with the HXR fluxes at 4-10 keV, 10-30 keV and 30-50 keV from RHESSI, and the EUV intensities at AIA 94, 131, 211, and 304Å .

Figure 3(a) A map of AR 11158 obtained by AIA 211 Å at UT 07:58 on February 15, 2011 is superimposed over HMI magnetogram. The magnetic field strength contours are divided in intervals of 95%, 80%, 60%, and 40% of the maximum values. The positive polarity is denoted by blue and the negative polarity in red. A clear jet feature appears on the left –hand side of the AR near the boundary.

(b) Time evolution of the EUV jet and the HXR source region of AR 11158 is in the time interval between 07:55 UT and 08:11 UT on February 15, 2011. RHESSI HXR images are reconstructed by the CLEAN algorithm using front segments of detectors 3 through 8. And the HXR contours are divided in intervals of 90%, 70%, and 50% of the maximum values. The green line is for 4-10 keV channel, blue for 10-30 keV and purple for 30-50 keV. The box with dashed lines outlines the area for summing count rates which are presented in Figure 4.

Figure 4  A integrated EUV intensity profile along the jet direction obtained by summing the count rates in the box as shown in the inset of Figure 3b. The images

from top to bottom: 94, 131, 211, 304 Å and with the Wind/Waves radio dynamic spectrum. The white line with the arrow head marks the loop top position.

Figure 5 Time evolution of the coronal temperature obtained from a differential emission measure (DEM) analysis of AR 11158 between 07:55 UT and 08:10 UT on February 15, 2011. The temperature range is indicated in the color bar on the right side of image (07:55 UT), log (T) = 5.7-7 (T~ 0.5-10 MK). The white contour marks the HXR sources in interval of 70%, 90% of maximum values. The black contour outlines the jet shape as observed in the associated 211 Å images.

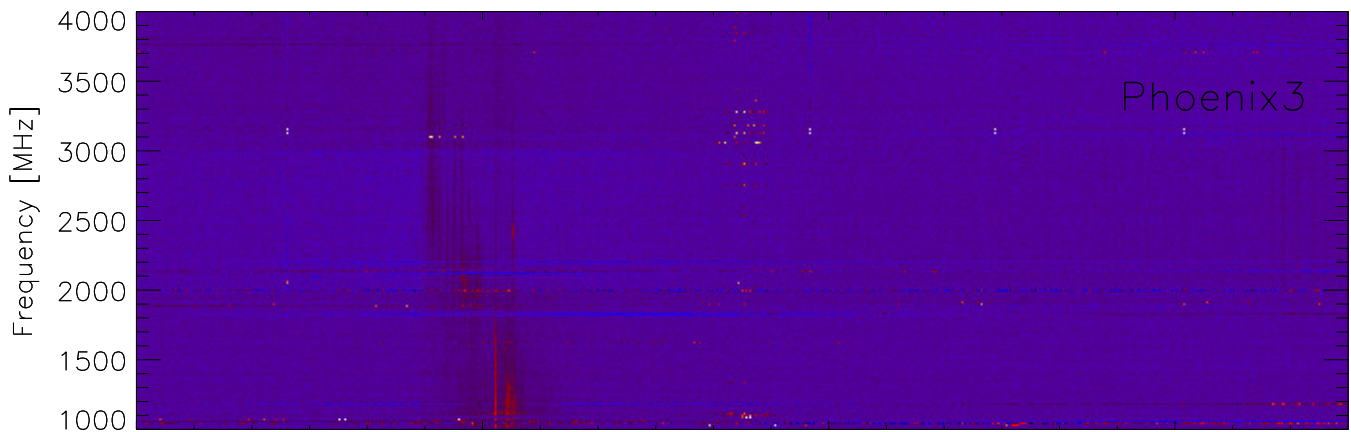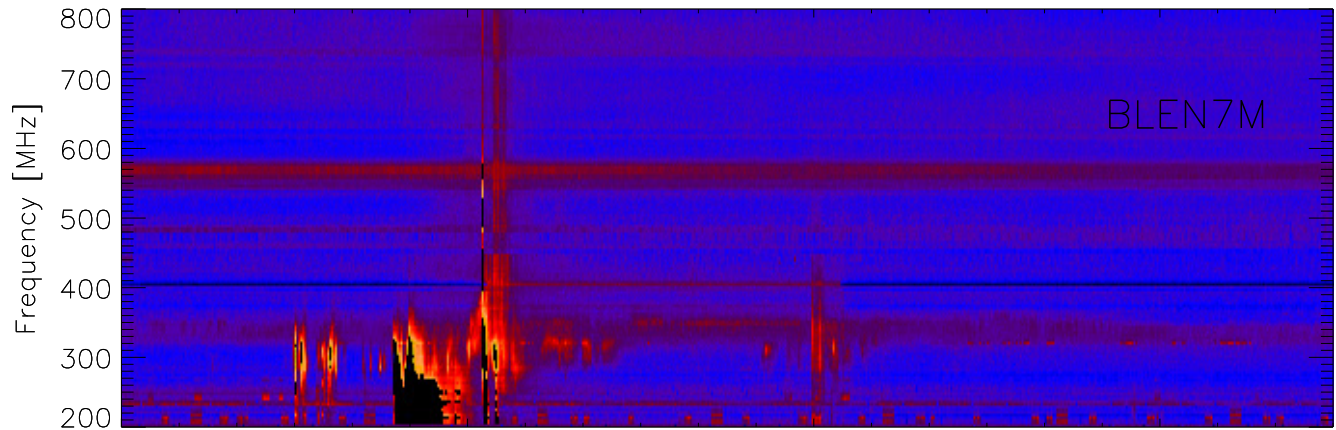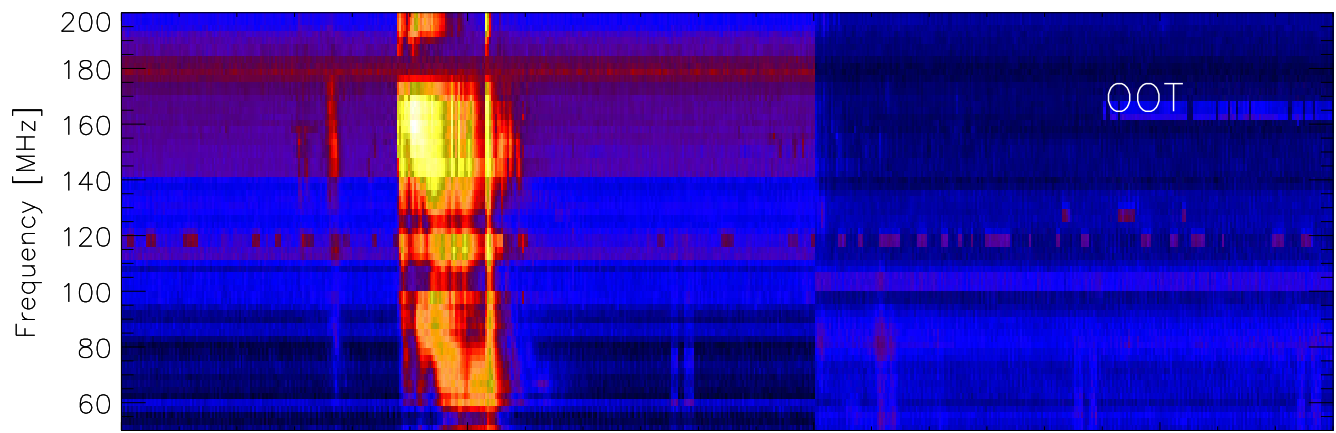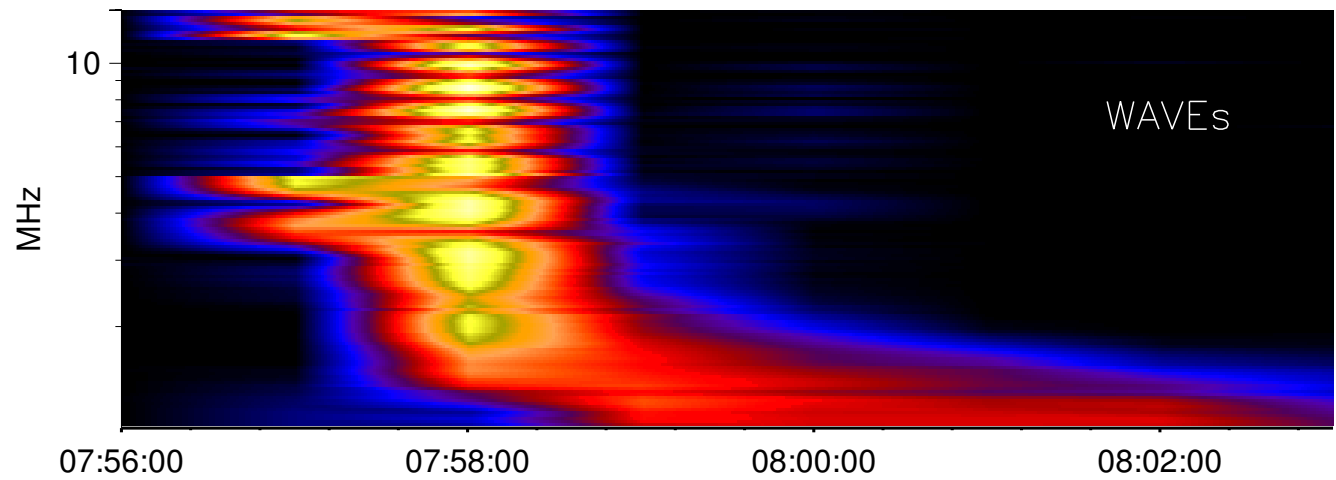

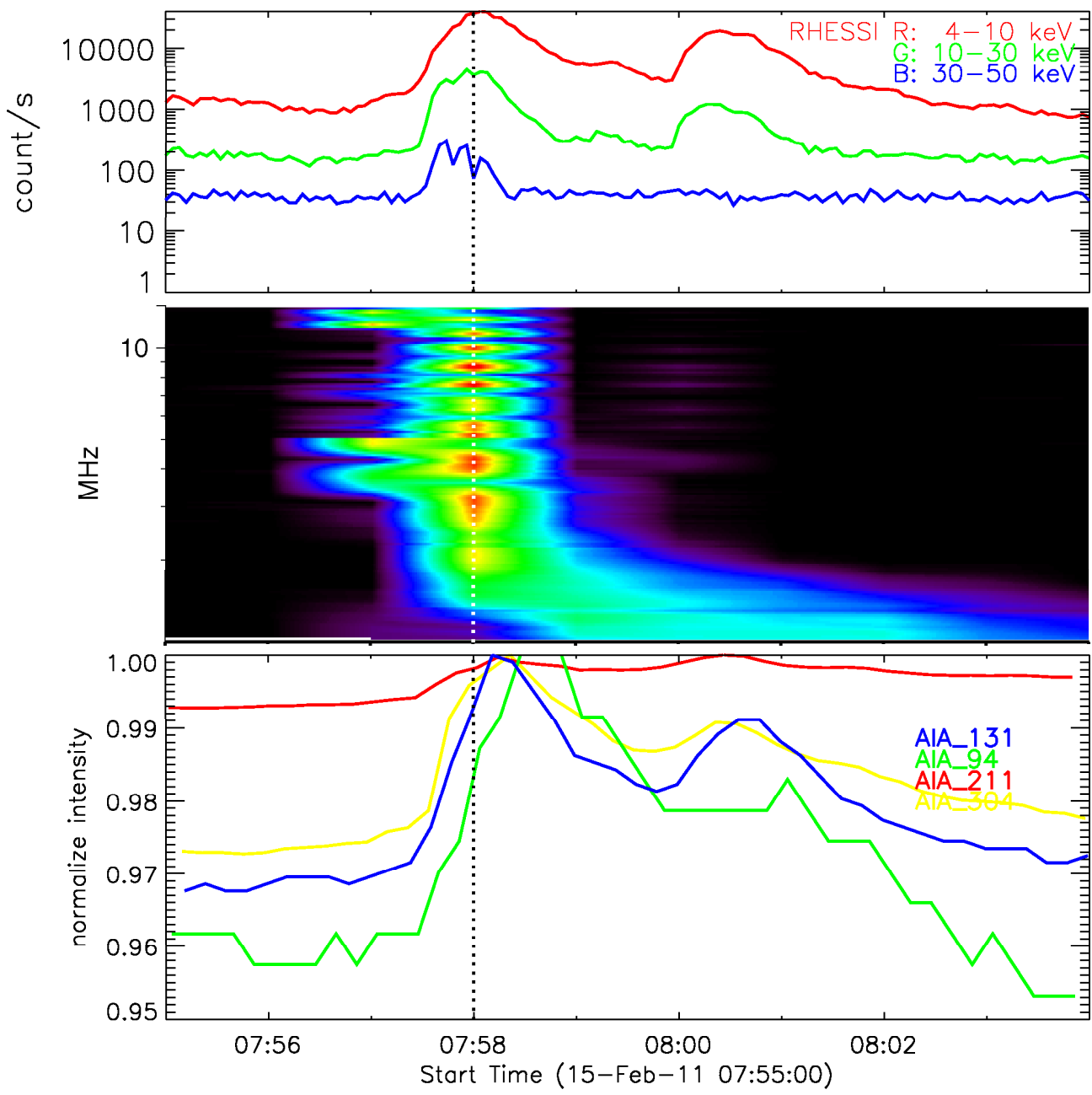

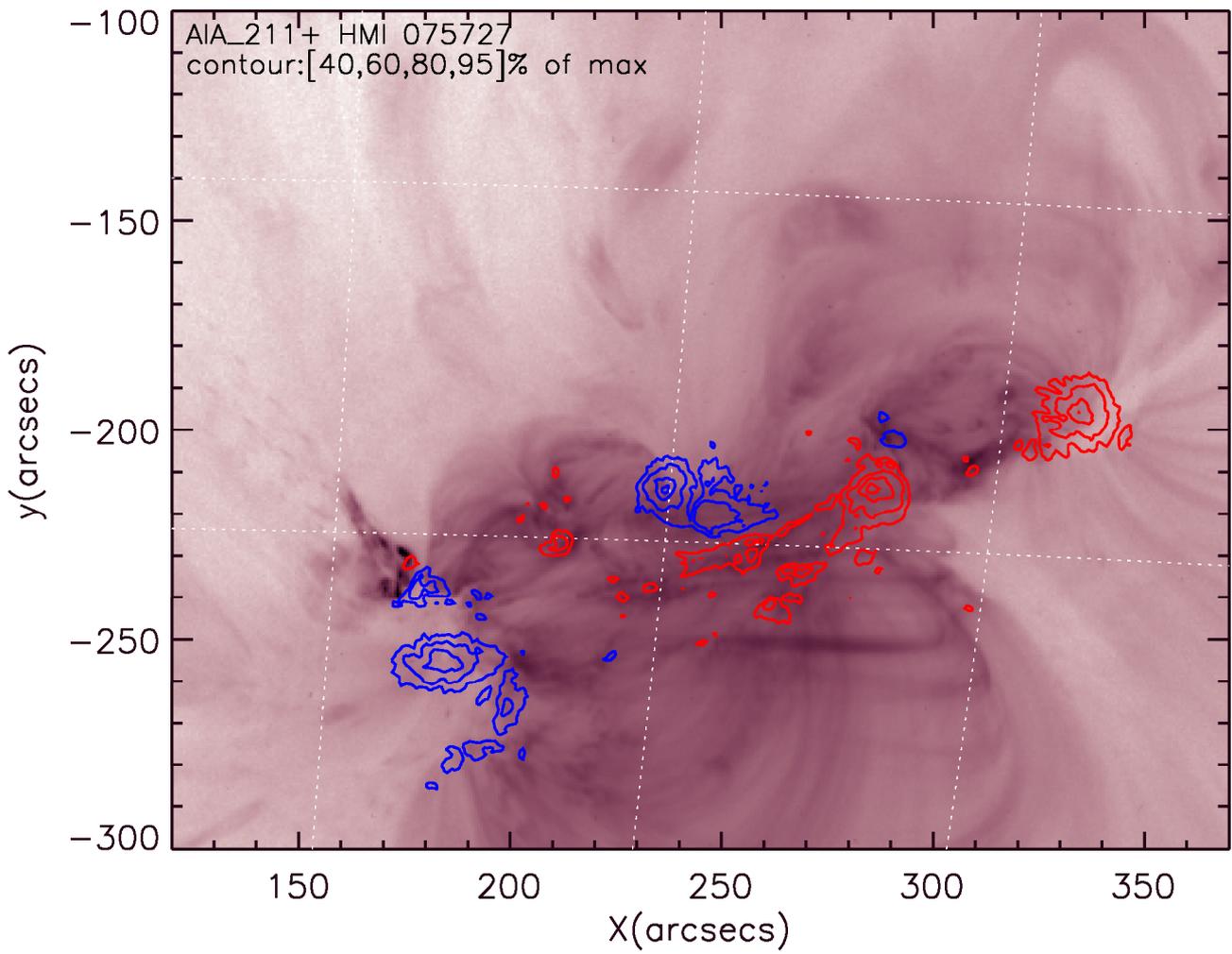

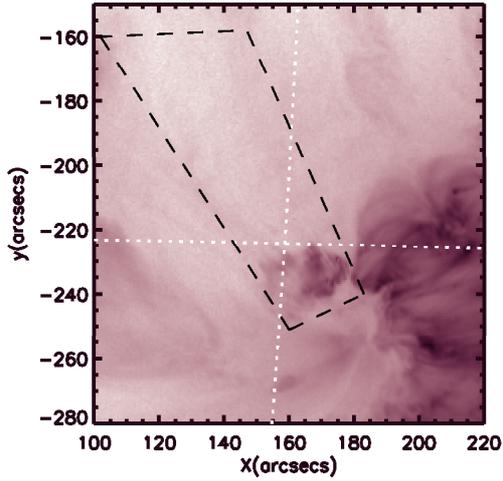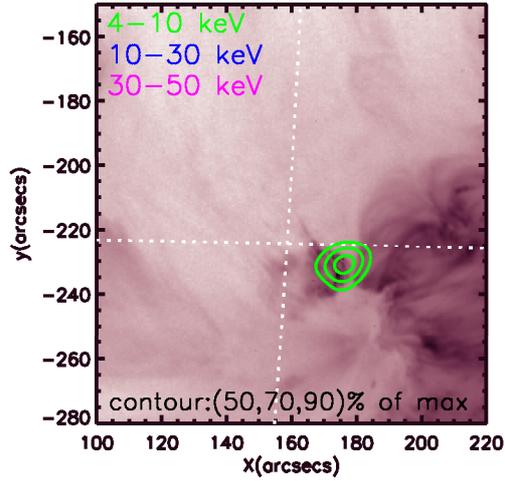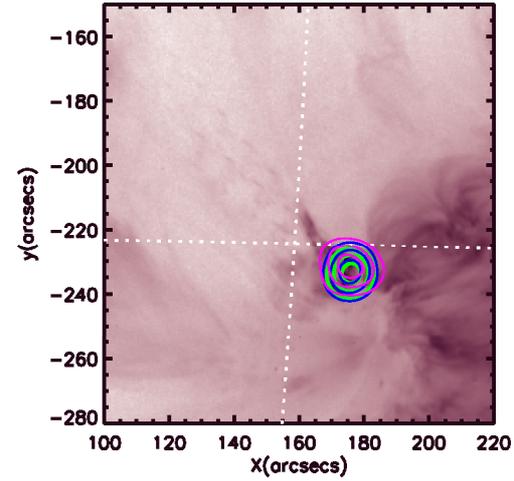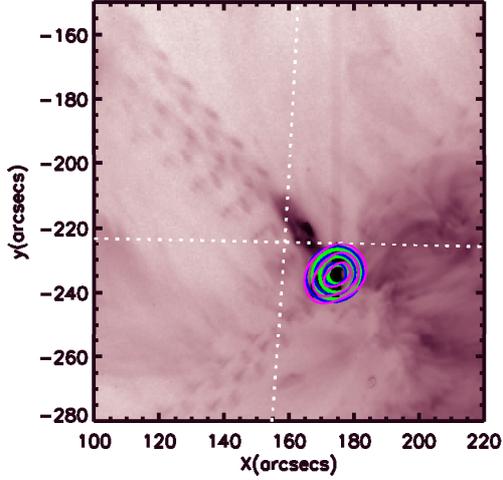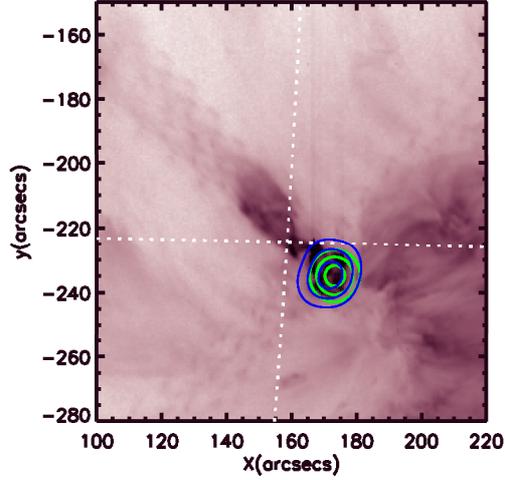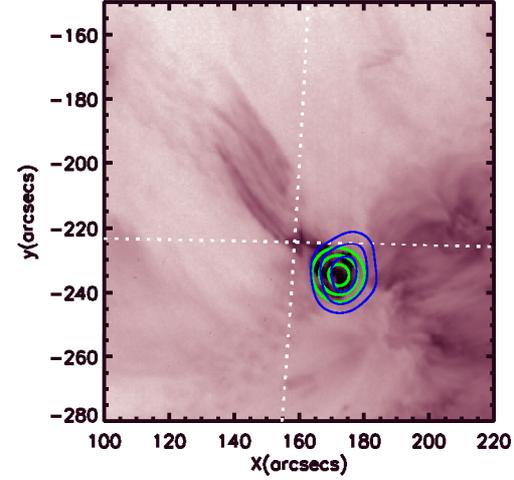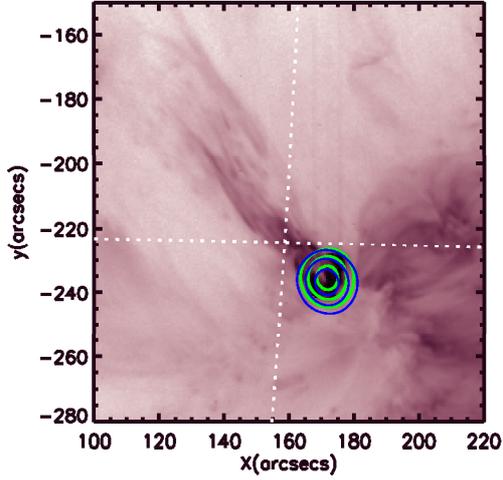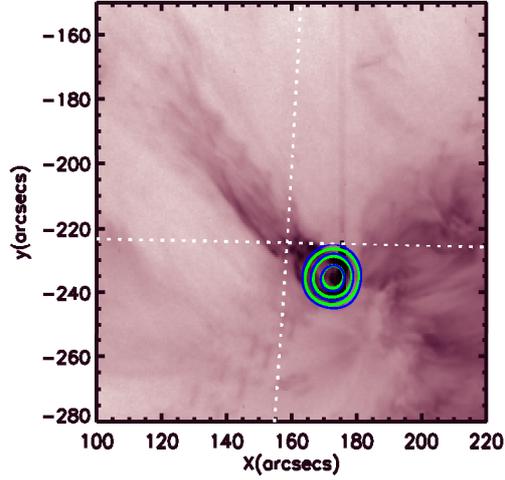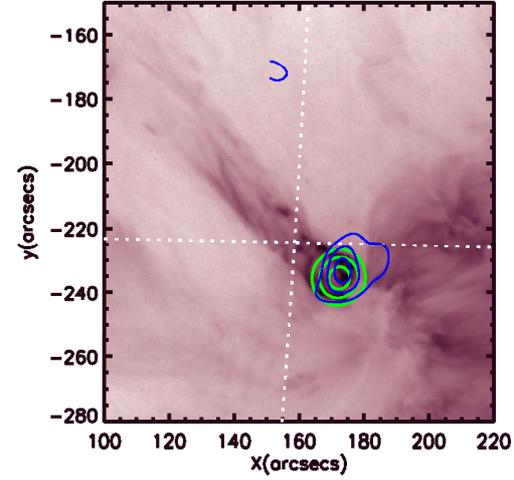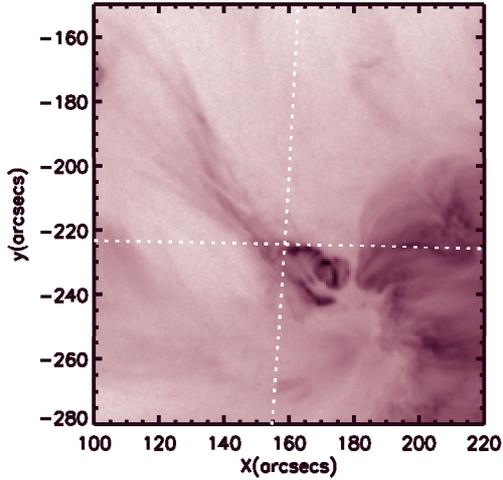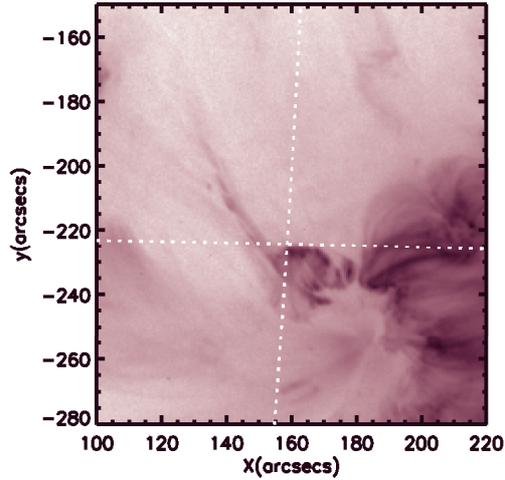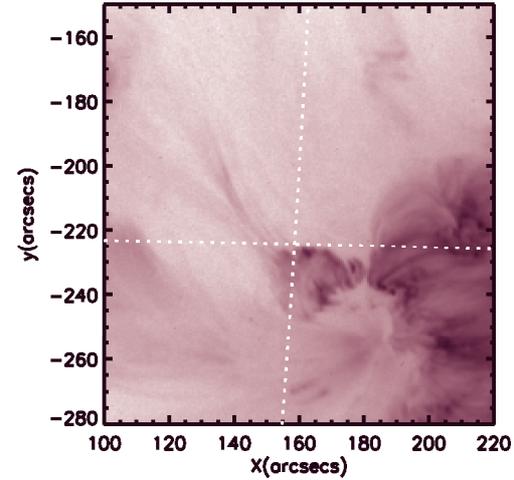

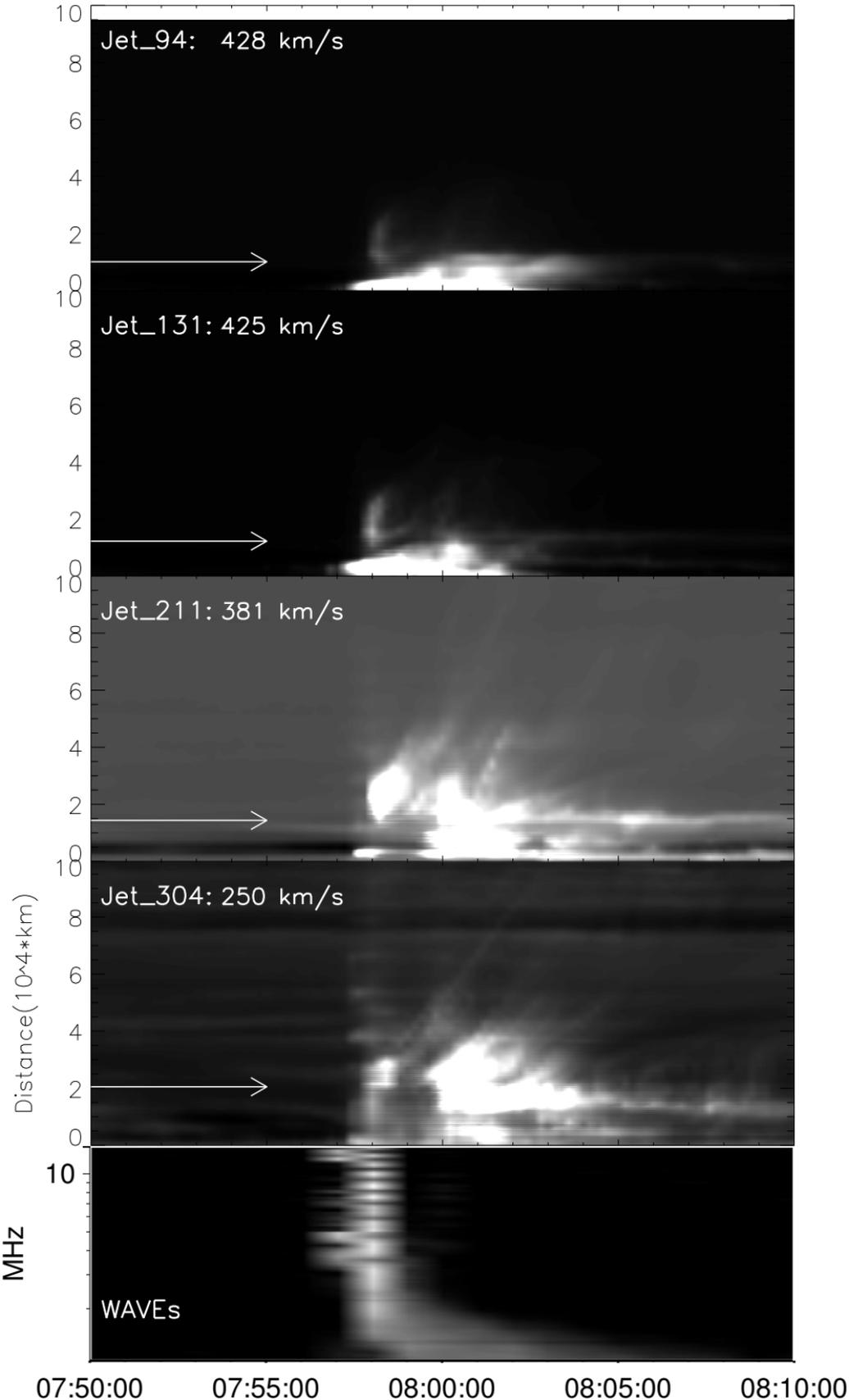

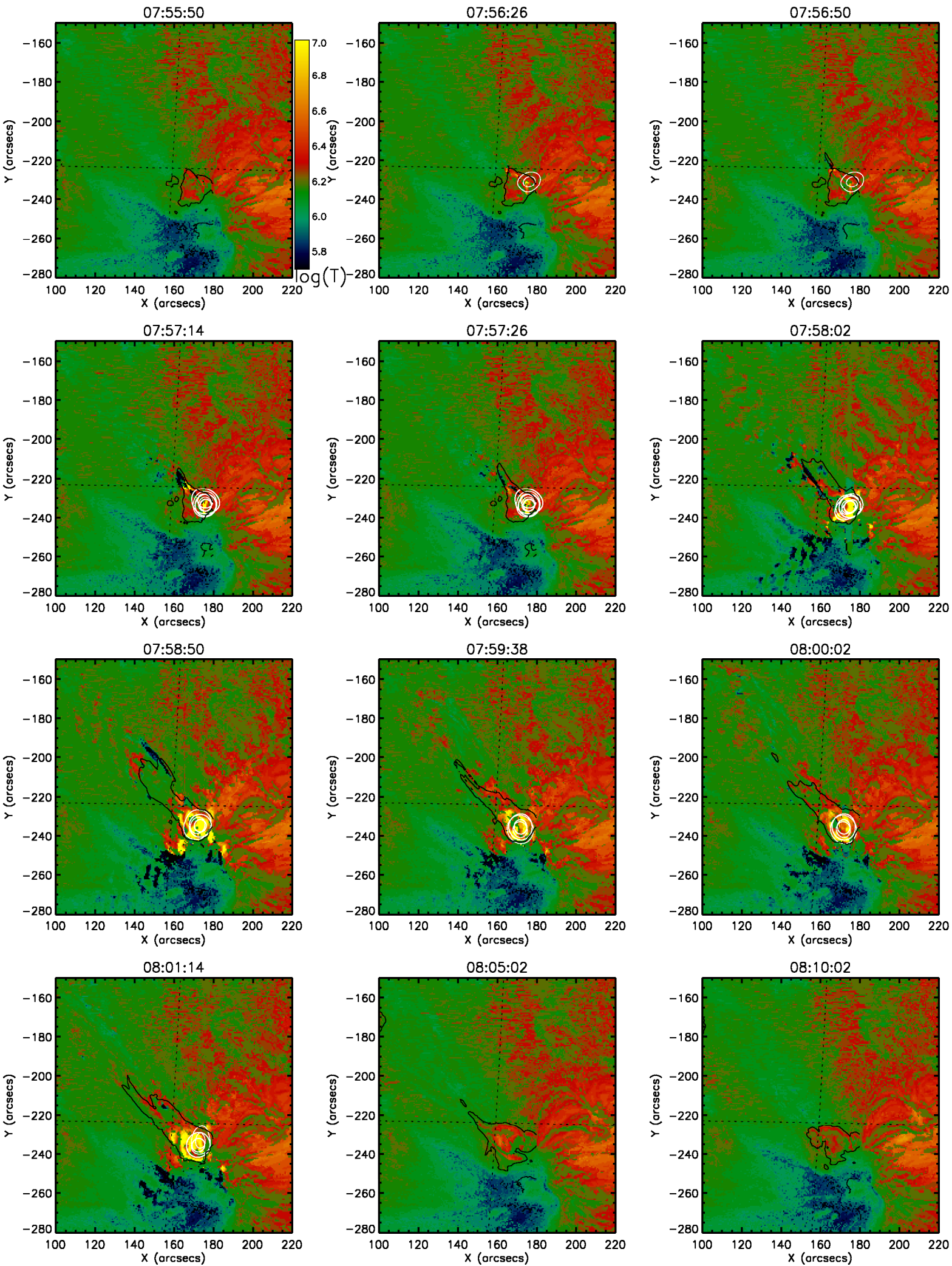